# ARTICLE

# Study of the electronic structure and optical absorption spectrum of the gold aromatic toroid $D_{6h} - Au_{42}$

Gennadiy Ivanovich Mironov*[a]

The electronic structure of the toroid $D_{6h} - Au_{42}$ is studied within the framework of the Hubbard Hamiltonian in the approximation of static fluctuations. An expression for the Fourier transform of the Green's function, the poles of which determine the energy spectrum of the nanocluster under consideration, is obtained. The energy spectrum of the $D_{6h} - Au_{42}$ toroid indicates that the nanosystem is in a metallic state. Graphical representations of the equation for the chemical potential and the density of state of electrons are presented. The spectrum of optical absorption is shown, the energies of direct optical transitions $D_{6h} - Au_{42}$ are 0.95 eV, 1.24 eV, 1.39 eV, are in the near infrared region. The energy of the ground state is calculated for the electrically neutral as well as for the positively and negatively charged ions of the toroid. The possibility of using of the investigated toroid from Au atoms for the diagnostics and treatment of cancer is shown.

## Introduction

Gold at the macrolevel and practically and at the microlevel behaves like a noble metal, basically does not interact with other substances, but at the nanoscale, gold begins to exhibit completely others, unique physicochemical properties. Nanoclusters consisting of a small number of gold atoms are in a semiconductor state, and only with an increase in the number of atoms, quantum systems of gold atoms begin to acquire features in the usual metallic state. The region of the nanoscale of gold near the semiconductor-metal transition has a number of interesting properties for practical application.

Much attention has been paid to studies of gold nanoclusters for more than two decades because of their unique catalytic [1–4], electronic [5–7], optical [8–11], and other properties. Studies of gold nanoclusters have led to the possibility of their practical application in the field of miniaturization of electronic devices [12], as building blocks in the synthesis of new functional materials [13], in medicine for the diagnostics and treatments of cancer [10, 11, 14-16]. In works devoted to the use of gold nanoparticles for diagnostics and for the treatment of cancer, the theoretical substantiation of the possible use of gold nanoparticles is either not given, or is based on the concept of plasmon resonance. It was shown in [10, 11] that the explanation of the properties of gold nanoparticles that can be used in cancer theranostics should not be based on the concept of surface plasmon resonance. As we will show below, the explanation of these properties should be based on the fact that the quantum system of d-electrons in gold nanoclusters is a system with strong correlations, and not an ordinary Fermi system.

By the time of publication of [17], only two hollow isomers of the $Au_{42}$ nanocluster have been depicted: the gold nanotube $D_{5d} - Au_{42}$ [18] and the icosahedral gold fullerene $I_h - Au_{42}$ [19]. Further exploration in [17] of the potential energy surface of hollow nanoclusters $Au_{42}$ led to the discovery of a toroidal analogue of a spherical gold cage – the gold aromatic toroid $D_{6h} - Au_{42}$ (Fig. 1).

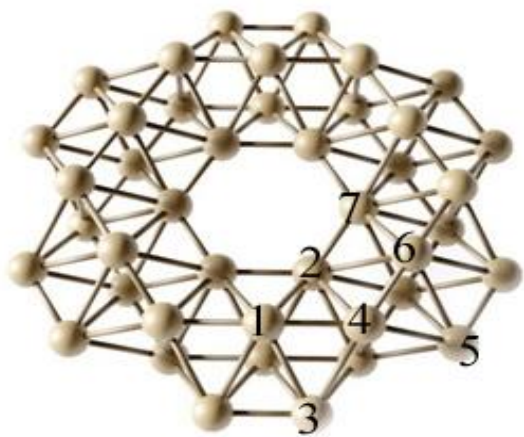

**Figure 1.** Structure of $D_{6h} - Au_{42}$ cluster

The structure $D_{6h} - Au_{42}$ consists of several octahedral subunits $Au_6$, the Au–Au distances range from 2.737 Å to 3.039 Å, they are similar to the Au–Au distances for isomers $D_{5d} - Au_{42}$ and $I_h - Au_{42}$ [17].

[a]     Prof. G.I. Mironov
Department of Physics and Materials Science
Mari State University
Kremlevskaya Street 44, Yoshkar-Ola (Russia)
° E-mail: mirgi@marsu.ru.

If in the case of the $D_{5d} - Au_{42}$ nanotube and the $I_h - Au_{42}$ fullerene each gold atom is surrounded by five or six atoms, then in the case of the $D_{6h} - Au_{42}$ toroid the gold atoms forming the $Au_6$ central pseudoring, interact with a large number of neighboring atoms. As a consequence, in [17] the gap width Δ between the highest occupied molecular orbital (HOMO) and the lowest unoccupied molecular orbital (LUMO) was 0.34 eV, which is less than the obtained value Δ = 0.505 eV for the $D_{5d} - Au_{42}$ nanotube [18] and the values Δ = 0.4 eV with BP86 and Δ = 0.9 eV B3LYP calculations for the $I_h - Au_{42}$ fullerene [19].

Calculations in [20-23] of the structure of the energy bands of gold nanoclusters showed that the energy levels of s-electron states in them are located completely within the valence band, they do not participate in electron transport. While the analysis of the density of d-electron states showed that the conduction band is formed by d-electrons, they are responsible for electron transport.

The aim of this work is to study the electronic structure of the $D_{6h} - Au_{42}$ toroid by the "quantum-field chemistry" method [7, 10, 11]. To study quantum systems in which d-electrons play the main role, the Hubbard model was proposed more than half a century ago [24, 25]. Within the framework of the Hubbard Hamiltonian, the metal atom is replaced by a model in which an electron, which determines the transport properties, moves around a positively charged ion. Depending on the degree of collectivization of these electrons, a metal-insulator phase transition is obtained [26]; the most significant properties of transition metals can be explained and the reasons for the peculiarities of the properties of gold nanoclusters can also be understood [7, 11].

## Theoretical model

The Hamiltonian of the Hubbard model as a system with strong correlations in the case of a toroid $D_{6h} - Au_{42}$ has the form:

$$\hat{H} = \varepsilon \sum_{\sigma} \sum_{f=1}^{42} \hat{n}_{f\sigma} + \sum_{\sigma, f \neq l} B_{fl}(a_{f\sigma}^+ a_{l\sigma} + a_{l\sigma}^+ a_{f\sigma}) + U \sum_{f=1}^{42} \hat{n}_{f\uparrow} \hat{n}_{f\downarrow}.$$

Where $a_{j\sigma}^+, a_{j\sigma}$ are Fermi-operators of creation and annihilation of $d-$electrons at the site $j$ ($j = f, l$) of the toroid with spin projection σ. $\hat{n}_{j\sigma} = a_{j\sigma}^+ a_{j\sigma}$ is the particle number operator, $\varepsilon$ is the energy of a d-electrons, $U$ is the Coulomb repulsion energy of $d-$electrons with oppositely oriented projections of spins in the same orbital, $B_{fl} = B(f - l)$ is the transfer integral. In the operator of total energy $\hat{H}$ the first term describes the self-energy of $d-$electrons, the second term describes the energy caused by the transfer of $d-$electrons from the toroid site to the neighboring site, the third term is the energy of the Coulomb repulsion of two $d-$electrons with different spin projections on the OZ axis, which find themselves at the same toroid site.

In [27, 28], a method for studying the Hubbard model in the "static fluctuation approximation" was proposed. In [29], within the framework of the static fluctuation approximation, the ground state energy of the one-dimensional Hubbard model was calculated for comparison with the exact solution of Lieb and Wu [30, 31]. It was shown in [29] that for the values of the Coulomb potential $U = 0$ and $U = \infty$, the approximate and exact solutions coincide; in the case of weak and strong couplings, the ground state energies of the one-dimensional model in the approximation of static fluctuations and in the case of the exact solution Lieb and Wu coincide with the good degree of accuracy, in the region of intermediate coupling, when the potential U is of the order of the width of the Hubbard subband, there is good agreement with the exact solution. The results obtained allowed us to conclude that in the approximation of static fluctuations it is possible to study systems described by the Hubbard model with a sufficient degree of accuracy, both in the region of weak and intermediate and strong correlations. The parameters of the Hubbard model for the aromatic toroid $D_{6h} - Au_{42}$ are such that the system of electrons describing the transport properties of the toroid is in the region of strong correlations. Later, using this method, it was possible to explain the observed optical absorption spectrum of fullerene C60 [10] and experiments on studying the energy spectra of "metallic" single-walled carbon nanotubes [27]. It turned out that in reality "metallic" single-wall carbon nanotubes are narrow-gap semiconductors.

## Results and Discussion

Having performed calculations in the framework of the static fluctuation approximation [2, 10], we obtain the following expression for the Fourier transform of the anticommutator Green's function in the case of the $D_{6h} - Au_{42}$ toroid:

$$\langle\langle a_1^+ | a_1 \rangle\rangle_E$$

$$= \frac{i}{2\pi} \sum_{n=1}^{2} \left\{ \frac{0.00727}{(E - \varepsilon_n + 0.105B} + \frac{0.00016}{E - \varepsilon_n - 1.678B} \right.$$

$$+ \frac{0.00491}{E - \varepsilon_n + 1.774B} + \frac{0.01378}{E - \varepsilon_n + 0.410B} + \frac{0.02121}{E - \varepsilon_n - 4.072B}$$

$$+ \frac{0.04591}{E - \varepsilon_n + 2.708B} + \frac{0.05322}{E - \varepsilon_n - 1.303B} + \frac{0.00120}{E - \varepsilon_n - 1.078B}$$

$$+ \frac{0.02589}{E - \varepsilon_n + 1.562B} + \frac{0.01487}{E - \varepsilon_n + 2.781B} + \frac{0.02405}{E - \varepsilon_n - 5.509B}$$

$$+ \frac{0.00907}{E - \varepsilon_n - 1.161B} + \frac{0.02431}{E - \varepsilon_n + 2.042B} + \frac{0.01212}{E - \varepsilon_n - 6.193B}$$

$$+ \frac{0.00339}{E - \varepsilon_n + 1.119B} + \frac{0.00203}{E - \varepsilon_n + 1.618B} + \frac{0.03011}{E - \varepsilon_n + 2.303B}$$

$$+ \frac{0.02599}{E - \varepsilon_n + 2.753B} + \frac{0.01578}{E - \varepsilon_n - 2.562B} + \frac{0.01578}{E - \varepsilon_n + 2.562B}$$

$$\left. + \frac{0.02589}{E - \varepsilon_n - 1.562B} + \frac{0.04167}{E - \varepsilon_n} + \frac{0.06301}{E - \varepsilon_n - 0.618B} \right\}$$

In the formula: $\varepsilon_n = \begin{cases} \varepsilon & , \quad n = 1 \\ \varepsilon + U & , \quad n = 2. \end{cases}$

The values of the numerator in fractions in the formula describe the probabilities of detecting an electron with a spin projection ↑ at the energy level determined when the denominator of this fraction is zero.

The poles of the Fourier transforms of the Green's functions describe the energy spectrum of electrons. In order to construct the energy spectrum of the toroid, it is necessary to determine the value of the electron's own energy $\varepsilon$ from the solution of the equation for the chemical potential for the toroid molecule $Au_{42}$.

For this, the Fourier transform of the Green's function is associated with the thermodynamic average $\langle \hat{n}_{1\uparrow} \rangle = \langle a_{1\uparrow}^+ a_{1\uparrow} \rangle$, which describes the average value of the number of electrons at site 1 with the spin projection ↑ [2]:

$\langle \hat{n}_{1\uparrow} \rangle$

$$= \sum_{n=1}^{2} \{ 0.00727 f^+(\varepsilon_n - 0.105B) + 0.00016 f^+(\varepsilon_n + 1.678B)$$
$$+ 0.00491 f^+(\varepsilon_n - 1.774B) + 0.01378 f^+(\varepsilon_n - 0.410B)$$
$$+ 0.02121 f^+(\varepsilon_n + 4.072B) + 0.04591 f^+(\varepsilon_n - 2.708B)$$
$$+ 0.05322 f^+(\varepsilon_n + 1.303B) + 0.00120 f^+(\varepsilon_n + 1.078B)$$
$$+ 0.02589 f^+(\varepsilon_n - 1.562B) + 0.01487 f^+(\varepsilon_n - 2.781B)$$
$$+ 0.02405 f^+(\varepsilon_n + 5.509B) + 0.00907 f^+(\varepsilon_n + 1.161B)$$
$$+ 0.02431 f^+(\varepsilon_n - 2.042B) + 0.01212 f^+(\varepsilon_n + 6.193B)$$
$$+ 0.00339 f^+(\varepsilon_n - 1.119B) + 0.00203 f^+(\varepsilon_n - 1.618B)$$
$$+ 0.03011 f^+(\varepsilon_n - 2.303B) + 0.02599 f^+(\varepsilon_n - 2.753B)$$
$$+ 0.01578 f^+(\varepsilon_n + 2.562B) + 0.01578 f^+(\varepsilon_n - 2.562B)$$
$$+ 0.02589 f^+(\varepsilon_n + 1.562B) + 0.04167 f^+(\varepsilon_n)$$
$$+ 0.06301 f^+(\varepsilon_n + 0.618B) \}.$$

In this formula, $f^+(x) = 1/((\exp(+x/kT) + 1))$ is the Fermi-Dirac distribution.

If we calculate in a similar way the average value of the number of electrons for all sites of the toroid, we obtain the formula for the average value of the total number of electrons $\sum_{i=1}^{42} (\langle n_{i\uparrow} \rangle + \langle n_{i\downarrow} \rangle)$ depending on the self-energy $\varepsilon$, i.e. equation for the chemical potential.

In fig. 2 is a graphical representation of the chemical potential equation. The ordinate is the number of electrons in the toroid, the abscissa is the self-energy in electron volts. In the case of an electrically neutral toroid, the eigenvalue $\varepsilon$ can have any value in the range from $-3\ eV$ to $-2.82\ eV$, usually the value in the center of the interval is chosen. In the case when one electron is missing, that is, in the case of the $Au_{42}^+$ ion, the value of its self-energy $\varepsilon = -2.8\ eV$. In the case of an electrically negatively charged ion $Au_{42}^-$ the self-energy value is $\varepsilon = -3.21\ eV$ in the center of the interval.

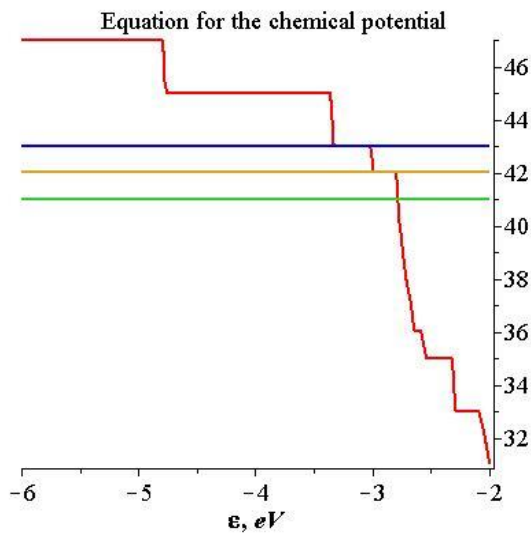

**Figure 2.** Graphical solution of the chemical potential equation

In fig. 3, the energy spectrum of the $D_{6h} - Au_{42}$ toroid is presented for the values of the model parameters $\varepsilon = -2.91\ eV, B = -1\ eV, U = 8.85\ eV$.

The self-energy $\varepsilon$ is determined from the equation for the chemical potential, the values for the electron transfer integral $B$ and the Coulomb potential $U$ are calculated by the optimization method from the known experimental values, for example, from gap width between the highest occupied molecular orbital and the lowest unoccupied molecular orbital in the case of hollow gold nanoclusters. If in the cases of the fullerene $I_h - Au_{42}$ and the nanotube $D_{5d} - Au_{42}$ the conduction band and the band of valence electrons do not overlap, they are in the semiconductor state, then in the case of the toroid $D_{6h} - Au_{42}$ these bands overlap. In fig. 3 inset demonstrates that above the lower level of the conduction band there are three upper levels of the valence electron band. The toroid is in a metallic state.

We noted above that in [17] the gap width between HOMO and LUMO in the case of the toroid $D_{6h} - Au_{42}$ is 0.34 eV. According to [17] the toroid is in the semiconductor state. The difference in the results is due to the fact that we took into account that the $D_{6h} - Au_{42}$ nanosystem is a system with strong correlations, while in [17] the toroid is considered a conventional Fermi system.

It is of interest to see how electron transfers occur from a toroid site to a neighboring site. For this, one can calculate thermodynamic averages characterizing the transitions of electrons from a site to a neighboring site, for example, the thermodynamic average $\langle a_{7\uparrow}^+ a_{2\uparrow} \rangle$, which describes the probability of an electron transition from the second site to the seventh site along the inner pseudoring of the toroid. In fig. 4, the form of this function is shown depending on the dimensionless quantity $x = -U/B$ for three cases: 1 - in the case of an electrically positively charged ion $Au_{42}^+$, 2 - in the case of an electrically neutral toroid $Au_{42}$, 3 - in the case $Au_{42}^-$.

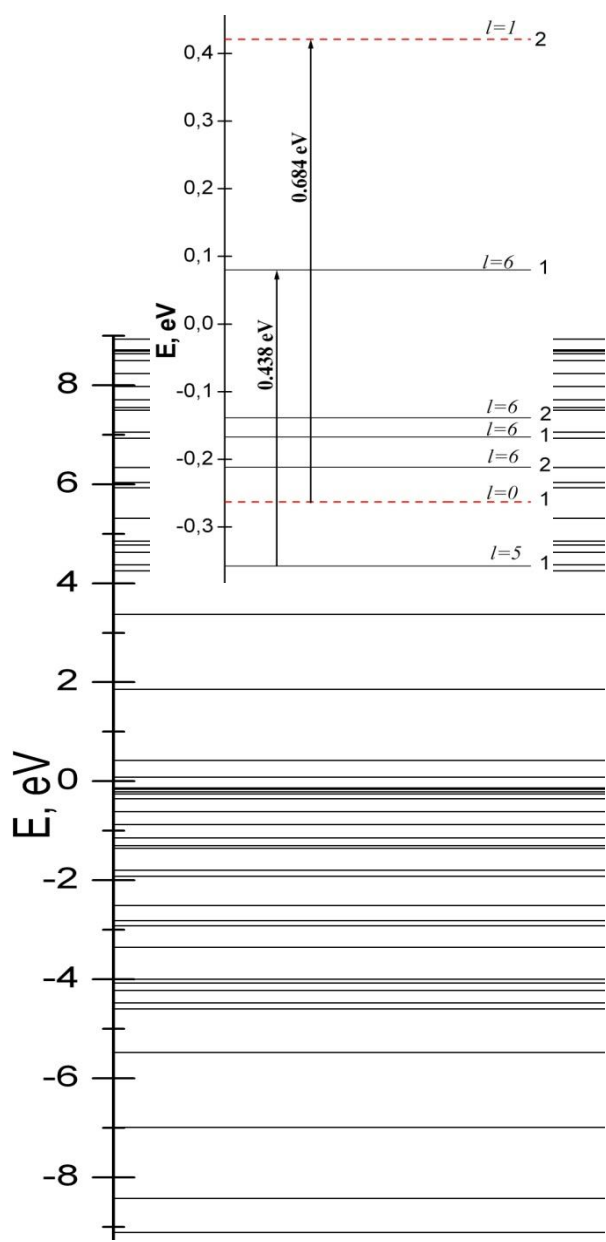

**Figure 3.** Energy spectrum of the $D_{6h} - Au_{42}$ toroid. Top panel: energy spectrum near 0 eV.

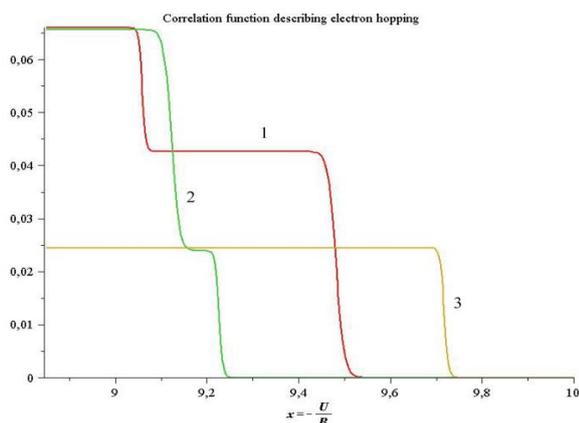

**Figure 4.** Thermodynamic average $\langle a_{7\uparrow}^+ a_{2\uparrow} \rangle$.

For the considered values of the parameters $B = -1 \, eV, U = 8.85 \, eV$, the probabilities of electron transfer along the inner pseudoring are practically the same in the cases of $Au_{42}^+$ and neutral $Au_{42}$, while the presence of an "extra" forty-third electron in $Au_{42}^-$ – more than halves the possibility of an electron transfer from site 2 to site 7 and in the opposite direction along the inner pseudoring. From the analysis of the graphs in fig. 4, it can also be concluded that in the case of a negatively charged ion of a toroid with a electron transfer integral $B = -1 \, eV$, electron jumps are possible even at a value of the Coulomb potential $U = 9.7 \, eV$. That is, despite the large value of the potential energy of repulsion of two electrons at one site, an electron can be transferred to a site where there is already an electron. Therefore, it is of interest to calculate the thermodynamic average characterizing the probability of finding two electrons with oppositely directed spin projections at one toroid site, for example, the thermodynamic average $\langle \hat{n}_{7\uparrow} \hat{n}_{7\downarrow} \rangle$. The form of the graphs in this case will correlate with the form of the graphs for the electron transitions in fig. 4.

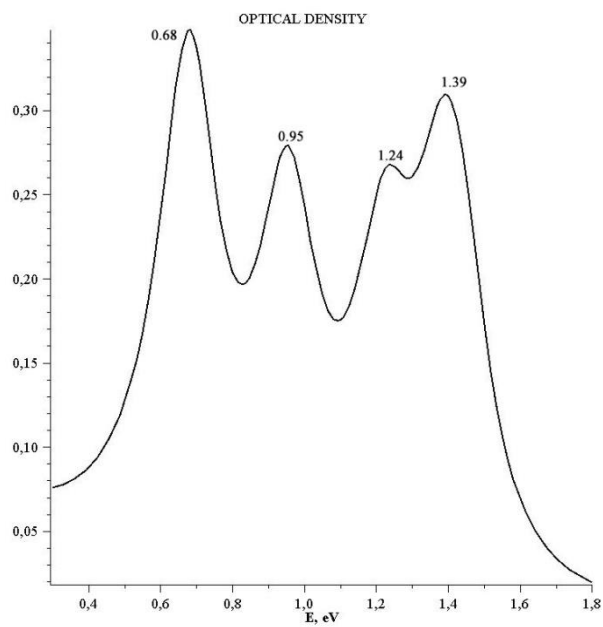

**Figure 5.** The optical absorption spectrum of the toroid at the following values of the parameters: $U = 8.85 \, eV, B = -1 \, eV, \varepsilon = -2.95 \, eV, C = 0.10 \, eV$.

Knowing the energy spectrum of the nanocluster, it is possible to determine the optical absorption spectrum. Fig. 5 shows the optical absorption spectrum of the toroid $D_{6h} - Au_{42}$ in the infrared region of the spectrum. The three peaks on the right side are in the near infrared region of the spectrum. It is interesting that, unlike the fullerene $I_h - Au_{42}$ and the nanotube $D_{5d} - Au_{42}$, the toroid $D_{6h} - Au_{42}$ is capable to absorb infrared radiation in an electrically neutral state. This is due to the fact that there is a slight overlap of the upper and lower Hubbard subzones, that the studied nanosystem is a metal.

Optical absorption in the near infrared region of the spectrum is of considerable applied interest. The fact is that radiation in this region passes through the soft tissues and blood of a living organism without strong absorption of radiation by them [33, 34]. Experiments indicate that hollow gold nanoclusters, when introduced into the body, accumulate on the surface of cancer cells [35-37]. If we now direct an infrared laser to the location of a cancerous tumor and toroids, then the waves waves corresponding to resonance peaks in the near infrared region will easily pass through soft tissues and blood vessels and reach the golden toroids. Optical absorption of infrared radiation will cause the appearance of ultrasonic and heat waves, recorded by a special sensor. As a result, a high-contrast image of a malignant tumor will appear on the screen of a photoacoustic tomograph [38, 39]. Therefore, with the help of toroids, it will be possible to produce photoacoustic visualization of a malignant neoplasm at an early stage of the disease.

It should be emphasized that the magnetic resonance imaging scans are limited to the number of resonating nuclei and can detect tumors of approximately 10 million tumor cells, which means that tumors are only detected when they reach a certain threshold. While photoacoustic imaging using gold hollow structures such as the toroid $D_{6h} - Au_{42}$ as contrast agents is a method that can help detect tumors at an earlier stage of cancer, which is very important since the development of malignant neoplasms is very rapid.

Thus, the investigated toroid can be used as a new class of materials for obtaining contrast improvements in the diagnostics of cancer at early stages, if an optimal solution for their synthesis is found. Once the diagnosis is confirmed, these same gold toroids can be used in cancer treatments. By directing an infrared laser to the location of a malignant neoplasm with toroids connected to its surface, we cause the absorption of infrared radiation by these hollow structures. As a result of nonradiative electron relaxation, a hollow gold nanocluster heats up and heats up tumor cells; at a local temperature above 43 ° C, cancer cells begin to die [39, 40]. Another option for cancer therapy is also possible. For this, let us first consider the energy of the ground state of the toroid $D_{6h} - Au_{42}$. The ground state energy is the average value of the Hamiltonian in the ground state:

$$\langle \bar{H} \rangle = \sum_{\sigma} \sum_{f=1}^{42} \varepsilon \langle \hat{n}_{f\sigma} \rangle$$

$$+ \sum_{\sigma, f \neq l} B_{fl} (\langle a_{f\sigma}^{+} a_{l\sigma} \rangle + \langle a_{l\sigma}^{+} a_{f\sigma} \rangle) + U \sum_{f=1}^{42} \langle \hat{n}_{f\uparrow} \hat{n}_{f\downarrow} \rangle.$$

Since we know all the thermodynamic averages in this formula, we can plot the dependence of $\langle \bar{H} \rangle$ on the dimensionless quantity $x = -U/B$. Figure 7 shows the energies of the ground state in three cases: 1 - in the case of a toroid ion $Au_{42}^{+}$, 2 - in the case of an electrically neutral toroid $Au_{42}$, 3 - in the case of an ion $Au_{42}^{-}$. In this figure, the ground state energy $\langle \bar{H} \rangle$ in electron volts is shown on the ordinate, and the abscissa is the dimensionless quantity $x = -U/B$.

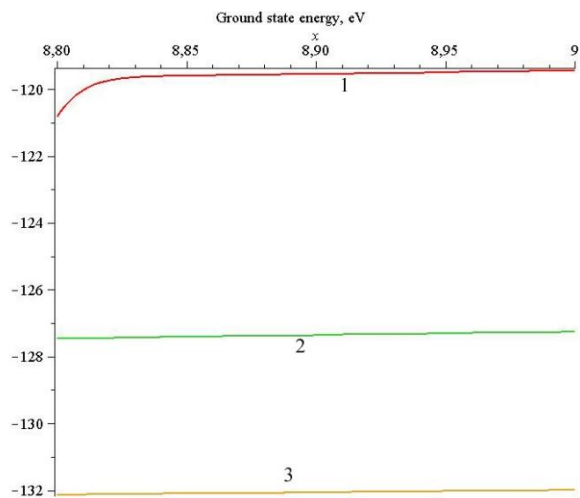

**Figure 6.** The energy of the ground state of the toroid as a function of the dimensionless parameter $x = -U/B$.

If, in the case of the values of the system parameters $B = -1 \ eV, U = 8.85 \ eV$, from the value of the energy of the ground state of the ion $Au_{42}^{+}$ subtract the value of the energy of the ground state of the electrically neutral toroid $Au_{42}$, we obtain the ionization energy of the toroid $Au_{42}$, it is equal to $E_I = 7.8 \ eV$. If we subtract the value of the energy of the ground state of the ion $Au_{42}^{-}$ from the value of the energy of the ground state of the electrically neutral toroid $Au_{42}$, we obtain the value of the energy of the electron affinity $E_A = 4.7 \ eV$. Thus, both the ionization energy and the electron affinity energy have positive values, comparable in order of magnitude with similar values, for example, in the case of carbon fullerene $C_{60}$ ($E_I = 7.6 \ eV$, $E_A = 2.7 \ eV$) [41, 42]. The combination of high electron affinity and lower ionization energy is a rare phenomenon in chemistry, indicating that toroids can be both donors and acceptors in chemical processes. This allows us to hope that the molecules of drugs for the treatment of cancer, target proteins for the search for malignant tumors will "stick" to the surface of the toroid, after which it is possible to carry out targeted delivery of drugs to the site of concentration of cancer cells [43-47]. If in [15] a synergistic effect was predicted when combining cancer treatment by hyperthermia and radiotherapy, then in our case, combining the action of drugs as a result of targeted delivery using hollow toroids and hyperthermia, we can assume that such a complex cancer treatment will give significant positive effect.

## Conclusions

Thus, within the framework of quantum field theory, it is possible to describe the properties of the toroid $D_{6h} - Au_{42}$. An analytical expression for the Fourier transform of the anticommutator Green's function is obtained describing the physicochemical properties of the gold hollow toroid. Thermodynamic averages describing the probabilities of an electron hopping from one site to another site of the toroid are defined. The energy of the ground state, which made it possible to determine the energy of affinity for the electron and the ionization energy of the toroid, is calculated. The spectrum of elementary excitations of the toroid indicates the presence of absorption of electromagnetic waves in the near infrared region of the spectrum. This property of the toroid can be used in cancer theranostics.

## Author Contributions

## Conflicts of interest

There are no conflicts to declare

## Acknowledgements

The acknowledgements come at the end of an article after the conclusions and before the notes and references.

## Notes and references

[1] M. Haruta, Nature 2005, 437, 1098-99.
[2] G.I. Mironov, The Physics of Metals and Metallography  2008, 105, 327-37.
[3] X. Zhang, S. Wang, Y. Liu, L. Li, C. Sun, APL Materials 2017, 5, 053501.
[4] X.J. Liu, I.P. Hamilton, Nanoscale 2017, 9, 10321-10326.
[5] P. N. D'yachkov, Russian Journal of  Inorganic Chemistry  2015, 60, 947-49.
[6] H. Ning, J. Wang, Q. Ma, H. Han, Y. Liu, Journal of  Physics and Chemistry of Solids 2014, 75, 596-99.
[7] G.I. Mironov, Russian Journal of  Inorganic Chemistry  2018, 63,  66-68.
[8] J. Chen, B. Wiley, Z. Li, D. Campbell, F. Saeki, H. Cang, L. Au, J. Lee, X. Li, Y. Xia, Advanced Materials 2005, 17, 2255-61.
[9] W.R. Hendren, A. Murphy, P. Evans, D. O'Connor, G.A. Wurtz, A.V. Zayats, R, Atkinson, R.J. Pollard, J. Phys.: Condens. Matter 2008, 20, 362203.
[10] G.I. Mironov, Physics of the Solid State 2019, 61, 1144-53.
[11] G.I. Mironov, Russian Journal of  Inorganic Chemistry  2019, 64, 1257-64.
[12] R.T. Senger, S. Dag, S. Ciraci, Physical Review Letters 2004, 93 (19), 196807.
[13] A.W. Castleman, S.N. Khanna, J. Phys. Chem. C 2009, 113, 2664-75.
[14] W.Li, P.K. Brown, L.V. Wang, Y. Xia, Contrast Media Mol. Imaging 2011, 6, 370-77.
[15] A. Zhang, W. Guo, Y. Qi, J. Wang, X. Ma, D. Yu, Nanoscale Research Letters 2016, 11, 279.
[16] S. Huang, C. Li, W. Wang, H. Li, Z. Sun, C. Song, B. Li, S. Duan, Y. Hu, International Journal of  Nanomedicine 2017, 12, 5163-76.
[17] A. Munos-Castro, ChemPhysChem 2016, 17,1-6.
[18] J. Wang, H. Ning, Q.-M. Ma, Y. Liu, Y.-C. Li, J. Chem. Phys. 2008, 129, 134705.
[19] Y. Gao, X.C. Zeng, J. Am. Chem. Soc. 2005,127, 3698-3699.
[20] E. P. D'yachkov, P. N. D'yachkov. J. Phys. Chem. C 2019, 123, 26005-10.
[21] P. N. D'yachkov. Russian Journal of Inorganic Chemistry 2020, 65,  1735-1738.
[22] P. N. D'yachkov, E. P. D'yachkov. Russian Journal of Inorganic Chemistry 2020, 65, 1196-1203.
[23] P. N. D'yachkov.Chemical Physics Letters 2020, 752, 137542.
[24] J. Hubbard,  Proc. Roy. Soc. A. 1963, 276, 238=257.
[25] J. Hubbard,  *Proc. Roy. Soc. A.* **1963**, *281*, 401-419.
[26] N. F. Mott, Metal - insulator transition,Taylor & Francis, London 1990.
[27] G. I. Mironov, Phys. Solid State 1997, 39, 1420-1424.
[28] G. I. Mironov, Phys. Solid State 1999, 41, 864-869..
[29] G. I. Mironov, Phys. Solid State 2002, 44, 216-220.
[30] E. H. Lieb and F. Y. Wu, Phys. Rev. Lett. 1968, 20, 1445-48.
[31] E. H. Lieb and F. Y. Wu, Physica A. 2003, 321, 1-27.
[32] G.I. Mironov, Low Temperature  Physics  2017, 43, 719-723.
[33] J. Lee, D.K. Chatterjee, M.H. Lee, S. Krishnan, Cancer Lett. 2014, 347, 46-53.
[34] J.F. Hainfeld, L. Lin, D.N. Slatkin, F.A. Dilmanian, T.M. Vadas, H.M. Smilowitz,  Nanomedicine 2014, 10, 1609-17.
[35] Y. Xia, W. Li, C. Cobley, J. Chen, X. Xia, Q. Zhang, M. Yang, E.C. Cho, P.B. Jarreau, Accounts of Chemical Research 2011, 44, 914-924.
[36] J. Cuo, K. Rahme, Y. He, L. Li, J.D. Holmes, C.M. O'Driscoll, Int. J. Nanomedicine 2017, 12, 6131-6152.
[37] B. Pang, X. Yang, Y. Xia, Nanomedicine 2016, 11, 1715-1728.
[38] G.I. Mironov, Phys. Solid State 2021, 63, 324-331.
[39] J.U.Menon, P. Jadeja, P. Tambe V. Khanh, Y. Baohong, K.T. Nguyen, Theranostics **2013**, *3*, 152-166.
[40]  P.Singh, S. Pandit, V.R.S.S. Mokkapati, A. Garg, V. Ravikumar, I. Mijakovic. Int. J. Mol. Sci. 2018, 19, 1979.
[41] R.K.Yoo, B. Ruscic, J. Bercowitz, J. Chem. Phys., **1992**, *96*, 911-918.
[42] C. Brink, L. Andersen, P. Hvelplund, D. Mathur, J.D. Volstad, Chem. Phys. Lett. **1995**, *233*, 52-56.
[43] G. Ajnai, A. Chiu, T. Kan, C. Cheng J. Exp. Clin. Med., **2014**, *6*, 172-178.
[44] Y.Tao, M. Li, J. Ren, X. Qu, Chem. Soc. Rev. **2015**, *44*, 8636-8663.
[45] L. Yang, Y. Tseng, G. Suo, L. Chen, J. Yu, W. Chiu, C. Huang, C. Lin, ACS Appl Mater Interfaces **2015**, *7*, 5097-5106.
[46] F.Y. Kong, J.W. Zhang, R.F. Li, Z.X. Wang, W.J. Wang, W. Wang, Molecules, **2017**, *22*, 1445.
[47] E.C. Dreaden, L.A. Austin, M.A.  Mackey,  M.A. El-Sayed,  Ther. Deliv. **2012**, *3*. 458-478.